\documentclass[12 pt]{article}
\pdfoutput=1
\usepackage{amsmath,amssymb,graphicx}
\usepackage{hyperref}
\usepackage[T1]{fontenc}
\usepackage[utopia]{mathdesign}
\usepackage[font=large,labelfont=bf]{caption}
\usepackage{longtable}
\usepackage{verbatim}
\usepackage{amsfonts}
\usepackage{cite}
\usepackage{enumerate}
\usepackage{url}

\usepackage[usenames,dvipsnames]{xcolor}
\definecolor{burgundy}{rgb}{0.565,0.0,0.125}

\usepackage{afterpage}

\newcommand{\bear}{\begin{array}}
\newcommand{\ear}{\end{array}}

\newcommand{\beq}{\begin{eqnarray}}
\newcommand{\eeq}{\end{eqnarray}}
\newcommand{\beqa}{\begin{eqnarray}}
\newcommand{\eeqa}{\end{eqnarray}}

\newcommand{\nn}{\nonumber}
\def\OMIT#1{{}}
\newcommand{\lsim}{\mathrel{\rlap{\lower4pt\hbox{\hskip1pt$\sim$}}
    \raise1pt\hbox{$<$}}}         
\newcommand{\gsim}{\mathrel{\rlap{\lower4pt\hbox{\hskip1pt$\sim$}}
    \raise1pt\hbox{$>$}}}         

\usepackage{titlesec}

\titleformat{\section}
{\color{burgundy}\normalfont\Large\bfseries}
{\color{burgundy}\thesection}{1em}{}

\titleformat{\subsection}
{\color{burgundy}\normalfont\large\bfseries}
{\color{burgundy}\thesubsection}{1em}{}

\title{\bf \color{burgundy} Heavy Gravitino and Split SUSY in the Light of BICEP2}

\author{JiJi Fan, Bithika Jain and Ogan \"Ozsoy \\
{\em Department of Physics, Syracuse University, Syracuse, NY 13244, USA}}

\begin{document}
\maketitle

\begin{abstract}
High-scale supersymmetry (SUSY) with a split spectrum has become increasingly interesting given the current experimental results. A SUSY scale above the weak scale could be naturally associated with a heavy unstable gravitino, whose decays populate the dark matter (DM) particles. In the mini-split scenario with gravitino at about the PeV scale and the lightest TeV scale neutralino being (a component of) DM, the requirement that the DM relic abundance resulting from gravitino decays does not overclose the Universe and satisfies the indirect detection constraints demand the reheating temperature to be below $10^9 - 10^{10}$ GeV. On the other hand, the BICEP2 result prefers a heavy inflaton with mass at around $10^{13}$ GeV and a reheating temperature at or above $10^9$ GeV with some general assumptions. The mild tension could be alleviated if SUSY scale is even higher with the gravitino mass above the PeV scale. Intriguingly, in no-scale supergravity, gravitinos could be very heavy at about $10^{13}$ GeV, the inflaton mass scale, while gauginos could still be light at the TeV scale. 
\end{abstract}

\section{Introduction}
\label{sec:intro}

Supersymmetry (SUSY) has long been a favorite theoretical framework of physics beyond the Standard Model (SM). However, given the current null results of all SUSY searches, if SUSY is realized in Nature, it is unclear at what scale it will manifest itself. At the moment, theoretical studies of SUSY fall into two broad catalogues: one direction is to still focus on weak-scale natural SUSY and design non-trivial structures of flavor and Higgs sectors to evade the direct search constraints and explain the observed Higgs mass. The other direction is take seriously high-scale fine-tuned SUSY, in particular, split SUSY, with scalars heavier than gauginos. The virtues of this approach include simplicity, automatic amelioration of SUSY flavor and CP problems, preservation of gauge coupling unification and the lightest neutralino being a dark matter (DM) candidate. The idea of split SUSY, in particular, mini-split with scalars one-loop factor heavier than gauginos, was actually predicted a while ago by the simplest version of anomaly mediation~\cite{Giudice:1998xp, Randall:1998uk} (and later by a wide variety of moduli mediation scenarios~\cite{Choi:2005ge,Choi:2005uz,Conlon:2006us,Conlon:2006wz,Acharya:2007rc,Acharya:2008zi}). Since 2003, split SUSY has started to be taken as a viable possibility despite the presence of a fine-tuned EWSB and gained more attention recently given the increasing tension between data and naturalness~\cite{Wells:2003tf, ArkaniHamed:2004fb,ArkaniHamed:2004yi,Giudice:2004tc,Acharya:2007rc,Acharya:2008zi,Hall:2011jd,Ibe:2011aa,Arvanitaki:2012ps,ArkaniHamed:2012gw,Hall:2012zp, Hall:2013eko, Altmannshofer:2013lfa, Baumgart:2014jya, Dhuria:2012bc}.

In split SUSY, the high SUSY breaking scale could naturally lead to a heavy unstable gravitino. In the mini-split scenario based on anomaly mediation, there is a loop factor separating the gravitino and gaugino mass scales with gravitino at about ($10^2 - 10^3$) TeV and gaugino at the TeV scale. In this scenario, the neutralino DM particles produced by late-time gravitino decays could not annihilate efficiently and thus inherit the number density of the gravitinos which adds to its thermal number density. During the reheating era, the thermal scattering of the SM superpartners contributes (at least part of) the gravitino primordial relic abundance, which is approximately proportional to the reheating temperature $T_R$. Consequently the requirement that the neutralino DM does not overclose the Universe sets an interesting upper bound on $T_R$ as a function of DM mass. This upper bound could be tightened if wino is (a component of) DM. Indirect detection looking for excesses in the photon continuum spectrum or a monochromatic photon line sets a strong bound on allowed wino DM relic abundance for the whole mass range assuming NFW or Einasto DM profiles~\cite{Cohen:2013ama, Fan:2013faa}. The bound could be relaxed if the Milky Way DM distribution near the galactic center deviates considerably from the standard DM-only $N$-body simulation predications. However, the bound does not necessarily disappear entirely. For example, even if the Milky Way DM profile has a significant core with a radius of 1 kpc, light non-thermal wino with mass below 400 GeV as a single-component DM is excluded~\cite{Fan:2013faa}. We will present the derivation of the upper bound on $T_R$ from the constraints of the relic abundance of neutralino DM, in particular, wino DM in Sec.~\ref{sec:relic} and Sec.~\ref{sec:indirect}. 

On the other hand, the discovery of $B$-mode by the BICEP2 collaboration gives us some clues of the inflation scale~\cite{Ade:2014xna}. The observation could be fit by a lensed $\Lambda$CDM plus tensor model with a tensor-to-scalar ratio $r = 0.2 ^{+0.07}_{-0.05}$. Such a large $r$ prefers large field inflation with a heavy inflaton and very likely a high reheating temperature. We will present estimates of inflaton mass scale and reheating temperature in Sec.~\ref{sec:bicep}. 

We find that in the mini-split scenario based on anomaly mediation, $T_R$ is bounded to be at or below $10^9 - 10^{10}$ GeV while the BICEP2 data prefers $T_R$ to be around or above $10^9$ GeV. The BICEP2 result has some tension with the mini-split scenario with a heavy gravitino. In other words, the BICEP2 result favors a splitting between gravitinos and gauginos larger than the loop factor predicted by anomaly mediation. Intriguingly, if SUSY breaking is tied up with gravity, e.g., through the Scherk-Schwarz mechanism, gravitinos could be as heavy as $10^{13}$ GeV, which is the same mass scale of the inflaton inferred from the BICEP2 result while gauginos could still be light at the TeV scale. The implications for SUSY scales will be discussed in Sec.~\ref{sec:implications}. See Refs.~\cite{Harigaya:2014qza, Pallis:2014dma, Harigaya:2014pqa, Ibanez:2014zsa, Craig:2014rta, Ellis:2014rxa, Lyth:2014yya, Hamaguchi:2014mza, Hall:2014vga} for some other recent discussions of implications  of the BICEP2 result for SUSY. 

We conclude in Sec.~\ref{sec:con} and present a discussion of gravitinos from inflaton decays in the appendix.

\section{Gravitino and Wino Relic Abundance}
\label{sec:relic}

In this section, we first review different mechanisms generating the primordial gravitino relic abundance in the early Universe. Then we discuss the relic abundance of wino DM from gravitino decays. Notice that most of the discussions also apply to other neutralino DM scenarios such as higgsino DM. The main point we want to emphasize is that: for gravitinos at or below the PeV scale, the neutralino DM relic abundance has an irreducible non-thermal contribution which scales linearly with the inflaton reheating temperature $T_R$; \textbf{in particular, requiring DM relic abundance not overclose the Universe restricts the reheating temperature to be below ($10^{10} - 10^9)$ GeV for DM mass in the range (100 GeV - 1 TeV)}; for gravitino much above the PeV scale, the neutralino DM relic abundance is almost UV insensitive, meaning that it is almost independent of $T_R$.

\subsection{Primordial Gravitino Relic Abundance}

As a superpartner of the graviton, the gravitino couples to all supermultiplets with gravitational interaction strength. In an $R$-parity conserving scenario, an unstable gravitino always decays to a particle and its superpartner. Decay of a gravitino will always produce a lightest superparticle (LSP) as all the other produced superparticles will cascade down to the LSP. The decay width of an unstable gravitino is given by 
\beq 
\Gamma_{3/2} \approx 2.0 \times 10^{-23}\, {\rm GeV} \left(\frac{N_G}{12} \right) \left(\frac{m_{3/2}}{100 \, {\rm TeV}}\right)^3,
\eeq
where $m_{3/2}$ is the gravitino mass and $N_G$ is the number of degrees of freedom gravitino decays to. In the split SUSY scenarios with all gauginos lighter than the gravitino and the squarks heavier than the gravitino, $N_G = 12$.\footnote{The squarks could be lighter than the gravitino in the split SUSY scenarios and then $N_G$ is larger. However, it will not change much our discussions and results.}

There could be several different origins of the primordial gravitino relic abundance, $\Omega_{3/2}h^2$. One comes from scattering processes of MSSM particles in the thermal bath~\cite{Pradler:2006hh, Rychkov:2007uq, Allahverdi:2004si}. This contribution approximately scales linearly with the inflaton reheating temperature $T_R$. The higher $T_R$ is, the larger the gravitino relic abundance is. We will use the following approximate formula for the gravitino yield:
\beq\label{gYUV}
Y_{3/2}^{UV} \approx \sum_{i=1}^3 y_i g_i^2(T_R) \ln \left(\frac{k_i}{g_i(T_R)}\right) \left(\frac{T_R}{10^{9}\,{\rm GeV}}\right),
\eeq 
where $y_{1,2,3} =  (0.653, 1.604, 4.276) \times 10^{-13}$,  $k_{1,2,3} = 1.266, 1.312, 1.271$ and $g_{1,2,3} (T_R)$ are gauge couplings of SM gauge group $U(1)_Y, SU(2)_W, SU(3)_c$ evaluated at $T_R$ respectively ~\cite{Pradler:2006hh}. The small $y$'s originate from $T_R/M_p$ with $M_p$ the reduced Planck scale. Compared to the formula given in~\cite{Pradler:2006hh}, we neglected a contribution at the order of $(M_i^2/m_{3/2}^2)$ with $M_i$ the gaugino masses. The yield given in \eqref{gYUV} leads to a gravitino relic abundance
\beq\label{OmUV}
\Omega_{3/2}^{UV}h^2\approx 5.1 \times 10^{-2} \left(\frac{m_{3/2}}{1\, \rm TeV}\right)\left(\frac{T_R}{10^9\, \rm GeV}\right),
\eeq 
where we evaluated temperature dependent variables in Eq. \eqref{gYUV} at $T_R=10^9\, \rm GeV$. In the numerical evaluation in Sec.~\ref{sec:indirect}, we include the full temperature dependence.

Another potential important contribution to the gravitino relic abundance comes from the decays of superpartners that are still in thermal equilibrium with the post-inflationary thermal bath~\cite{Cheung:2011nn}.\footnote{This contribution exists only when $m_s > m_{3/2}$, where $m_s$ denotes the SUSY scalar mass. Besides, the thermal equilibrium requires reheating temperature to be $T_R>m_s$.} 
When the temperature of the primordial plasma drops around the SUSY scalar masses, which we will take to be around the same scale, decays of the scalars to gravitinos could also generate a potentially non-negligible contribution to the gravitino relic abundance. This is the ``freeze-in" mechanism~\cite{Hall:2009bx}. 
When the temperature drops below the scalar masses, the number density of SUSY scalars is suppressed exponentially, $e^{-m_s/T}$ and freeze-in stops. The freeze-in contribution is independent of the UV physics, particularly the reheating temperature $T_R$~\cite{Cheung:2010gj}. The gravitino yield from freeze-in is
\beq\label{gYFI}
\nn Y_{3/2}^{FI}&\simeq & \frac{405}{4\pi^4}\sqrt{\frac{5}{2}}\frac{M_{p}}{g_*^{3/2}} \sum_i g_i\frac{\Gamma_i}{m_i^2},\\ 
&\approx &1.6\times 10^{-16} \left(\frac{200}{g_*}\right)^{3/2} \left(\frac{100\, \rm TeV}{m_{3/2}}\right)^2\sum_{i} g_i \left(\frac{m_i}{1000\, \rm TeV}\right)^{3},
\eeq
where we approximated $g_*(m_i)\simeq g_{S*}(m_i)$ and $\Gamma_i=(1/48\pi)(m_i^5/(m_{3/2}^2M_{p}^2))$ as the partial decay width of scalar $i$ to the gravitino. Here, $g_i$ denotes degrees of freedom of SUSY scalar $i$ with mass $m_i$. The yield in Eq. \eqref{gYFI} leads to a gravitino relic abundance
\beq\label{OmFI}
\Omega_{3/2}^{FI}h^2\approx 1.1 \times 10^{-2} \left(\frac{100\, \rm TeV}{m_{3/2}}\right)\sum_{i} g_i \left(\frac{m_i}{1000\, \rm TeV}\right)^{3}.
\eeq 
It is clear that the gravitino relic abundance $\Omega_{3/2}h^2$ from the freeze-in contribution is highly sensitive to the scalar superpartner masses $m_i$ as it scales as $\sim m_i^3$.  
  
The total gravitino abundance is just a sum of the thermal scattering (Eq. \eqref{OmUV}) and freeze-in (Eq. \eqref{OmFI}) contributions
\beq
\Omega_{3/2}h^2=\Omega_{3/2}^{UV}h^2+\Omega_{3/2}^{FI}h^2.
\eeq

Before ending this section, we want to mention that there could be other model-dependent sources of primordial gravitino relic abundance. For example, decay of inflaton itself could also produce a sizable gravitino relic abundance. The contribution to gravitino relic abundance from inflaton decays depends on the structure of the dynamical SUSY breaking sector and could be problematic~\cite{Kawasaki:2006gs,Kawasaki:2006hm}. However, as discussed in~\cite{Dine:2006ii, Nakayama:2012hy}, gravitino production from inflaton decay can be suppressed if there exists a hierarchy between the mass scales of the inflaton and the field whose $F$-term VEV breaks SUSY spontaneously. In the discussions below, we will not include this model dependent contribution. We refer the reader to the Appendix A for more details of gravitinos from inflaton decays.

\subsection{Wino Relic Abundance from Gravitino Decays}
In this section, we will specify the neutralino DM to be wino yet the discussions hold for other neutralino DM such as higgsino DM. We will also focus on gravitino with mass above 10 TeV so that its lifetime is shorter than a second and its decays do not spoil the successful Big Bang Nucleosynthesis~\cite{Kawasaki:2008qe}. 

The relic abundance of wino DM is a sum of the thermal contribution and the non-thermal contribution from gravitino decays.
The non-thermal contribution could be computed numerically by solving the Boltzmann equations 
Eq. (2.1) - (2.3) in Ref~\cite{Moroi:2013sla}. The primordial gravitino relic abundance in Eq. \eqref{gYUV} and \eqref{gYFI} discussed in the previous section is an input to the Boltzmann equations. In solving the Boltzmann equations, we took $g_*(T)$ and $g_{*,s}(T)$ from a table in the DarkSUSY code~\cite{Gondolo:2004sc}.\footnote{We keep factors involving $\partial\log g_{*(s)}(T)/\partial\log T$ in the Boltzmann equation for $\rho_{\rm rad}$.} As the Sommerfeld effect becomes important for heavy winos~\cite{Hisano:2006nn,Hryczuk:2011vi,Moroi:2013sla}, we computed the temperature-dependent value of $\left<\sigma_{\rm eff} v\right>$ from a preliminary version 1.1 of the DarkSE code~\cite{Hryczuk:2011tq}, taking into account not only the Sommerfeld effect but also co-annihilation among different wino species.\footnote{This version was kindly provided by Andrzej Hryczuk to JF in a previous project.} As an input to this code, we have used the two-loop splitting 
between the neutral and charged winos from Ref.~\cite{Ibe:2012sx}. For wino masses of about a TeV and temperatures around a GeV, the Sommerfeld enhancement can be as large as 3 in $\left<\sigma_{\rm eff} v\right>$. 

The non-thermal contribution to wino relic abundance from gravitino decays changes parametrically when the gravitino mass $m_{3/2}$ increases. For large gravitino mass, the wino LSP produced from the gravitino decays can annihilate effectively due to the high temperature of the plasma at the time of gravitino decay. More specifically, we find that DM annihilation is efficient for $m_{3/2} \gtrsim10^4$ TeV. This can be seen by estimating the ``decay temperature" as in~\cite{Moroi:2013sla} 
\beq\label{DTg}
T_{3/2} \equiv \left(\frac{10}{g_*(T_{3/2}) \pi^2}M_{\rm pl}^2 \Gamma_{3/2}^2\right)^{1/4} \approx 2.2\,{\rm GeV}\left(\frac{75.75}{g_*(T_{3/2})}\right)^{1/4}\sqrt{\frac{N_G}{12}} \left(\frac{m_{3/2}}{10^4\, {\rm TeV}}\right)^{3/2}
\eeq    
At such high temperature, winos produced from the gravitino decays annihilate rapidly, reducing the number density $n_{\tilde{W}}$ down to a critical value $n_{c,\tilde{W}}\simeq 3H/\left<\sigma_{\rm eff} v\right> |_{T=T_{3/2}}$ at which winos can no longer annihilate. This critical value $n_{c,\tilde{W}}$ behaves as an attractor in determining relic abundance of wino LSP, making it independent of the primordial gravitino relic abundance. In this case, the wino relic abundance is given as 
\beq\label{ann-OmW}
\Omega_{\tilde{W}}^{(ann)}h^2&\approx &  m_{\tilde{W}}\frac{3H}{\left<\sigma_{\rm eff} v\right> s} \left |_{T=T_{3/2}}\right .\left( \frac{h^2}{\rho_{c,0}/s_0}\right) ,\\
&\approx & 0.12~\left(\frac{75.75}{g_*(T_{3/2})}\right)^{1/4} \left(\frac{m_{\tilde{W}}}{1~\rm TeV}\right)\left(\frac{1.2\times 10^{-7}~{\rm GeV}^{-2}}{\left<\sigma_{\rm eff} v\right>(T_{3/2})}\right)\left(\frac{m_{3/2}}{10^4 {\rm TeV}}\right)^{-3/2}
\eeq
where we used Hubble parameter $H(T)=\sqrt{g_*(T)\pi^2/90}T^2/M_{p}$, entropy density $s(T)=2\pi^2 g_{*,s}(T)T^3/45$. We also assumed $g_*\simeq g_{s,*}$ for the temperature of interest. We will present a more precise numerical evaluation in the following section. 

For a lighter gravitino within the mass range, $10~{\rm TeV} < m_{3/2} < 10^4~{\rm TeV}$, the gravitino starts to decay at such a low temperature that the annihilation of wino DM is ineffective. 
In this case, almost all the winos produced from gravitino decays survive and hence, its relic abundance is proportional to the total gravitino abundance.
\beq\label{no-annOmW}
\Omega_{{\tilde{W}}}^{(no-ann)} h^2&=&\frac{m_{\tilde{W}}}{m_{3/2}}\left(\Omega_{3/2}^{UV}h^2+\Omega_{3/2}^{FI}h^2\right) \nonumber \\
& \approx & 0.12 \left(\frac{m_{\tilde{W}}}{1 \, {\rm TeV}}\right)\left[\left(\frac{T_R}{2\times 10^9 \, {\rm GeV}}\right)+ 10^{-3}\left(\frac{100 \, {\rm TeV}}{m_{3/2}}\right)^2\sum_{i} g_i \left(\frac{m_i}{1000~ \rm TeV}\right)^{3}\right], 
\eeq
where the first(second) term in the square brackets in the second line originates from decays of gravitino produced by the thermal scattering (freeze-in). We want to caution the reader that there is no sharp boundary value of $m_{3/2}$ that separates the two cases with ``effective'' and ``ineffective'' wino annihilations in Eq.~(\ref{ann-OmW}) and Eq.~(\ref{no-annOmW}). In Sec.~\ref{sec:indirect}, we will derive more precise bounds by solving the Boltzmann equations numerically. 

From Eq.~(\ref{no-annOmW}), we could see that for gravitino at or below PeV scale as in the mini-split scenario, to avoid overproduction of DM from gravitino decays, the reheating temperature has to be below
\beq \label{eq:trupper}
T_R \lesssim 2 \times 10^9\,{\rm GeV} \left(\frac{1\,{\rm TeV}}{m_{\tilde{W}}}\right),
\eeq
assuming a negligible contribution from freeze-in. This upper bound would only be pushed even lower if the freeze-in contribution is comparable to or even dominate over the thermal scattering contribution. Similarly, one could obtain an upper bound on the scalar soft mass
\beq
m_s \lesssim 10^4\,{\rm TeV} \left(\frac{m_{3/2}}{100\,{\rm TeV}}\right)^{2/3} \left(\frac{1\,{\rm TeV}}{m_{\tilde{W}}}\right)^{1/3}. 
\eeq
Early discussions of reheating temperature in high-scale SUSY scenario with a decaying gravitino could be found in~\cite{Kallosh:2011qk, Dudas:2012wi}.

\section{Indirect Detection Constraints}
\label{sec:indirect}
As wino DM has a large annihilation rate, there are strong constraints on its relic abundance from indirect detection searches looking for its annihilation products~\cite{Ackermann:2011wa,Ackermann:2012kna,Ackermann:2013uma,Abramowski:2011hc,Alex:FermiLAT}. Thus in the wino DM case, one could obtain a stronger upper bound on the reheating temperature compared to Eq.~(\ref{eq:trupper}) which holds for generic neutralino DM. In this section, we present a numerical evaluation of the constraints on the reheating temperature and SUSY scalar mass scale in the scenario with wino as (a component of) DM. 

There are multiple indirect search channels for wino DM~\cite{Hryczuk:2014hpa}. In general DM indirect detection searches for decay and annihilation products of DM in fluxes of cosmic rays containing charged particles or photons or neutrinos. We focus on searches looking for excesses in the photon continuum spectrum of satellite dwarf galaxies ~\cite{Ackermann:2011wa,Alex:FermiLAT}, or our galactic center ~\cite{Hooper:2012sr} and monochromatic photon line~\cite{Abramowski:2013ax, Ackermann:2013uma}.\footnote{The first paper on the HESS search constraint for wino DM is Ref.~\cite{Cirelli:2007xd}.} A continuum photon spectrum is generated from either the bremsstrahlung of charged particles or the hadronic fragmentation of the decay products of $W/Z$'s in the final state of tree-level processes $\chi^0 \chi^0 \rightarrow W^{+} W^{-}/ZZ$. The gamma ray lines are generated from DM annihilation into $\gamma \gamma / \gamma Z$. Each photon in the final state carries away energy about the DM mass.

As demonstrated by Fig. 4 in Ref.~\cite{Fan:2013faa}, the thermal wino relic abundance (computed in~\cite{Hisano:2006nn, Cirelli:2007xd}) is ruled out by the indirect constraint for $m_{\tilde{W}}$ above 1.5 TeV assuming standard cuspy (NFW and Einasto) DM halo profiles. Since the wino relic abundance is a sum of the thermal contribution and the non-thermal contribution from gravitino decays, there is room for a non-thermal relic abundance only for wino with mass below 1.5 TeV.\footnote{There could be different non-thermal scenario such as moduli scenario~\cite{Moroi:1999zb}. The implications of indirect detection for moduli scenario have been discussed in~\cite{Fan:2013faa, Easther:2013nga, Allahverdi:2013noa}.}

\begin{figure}[!h]\begin{center}
\includegraphics[width=0.5\textwidth]{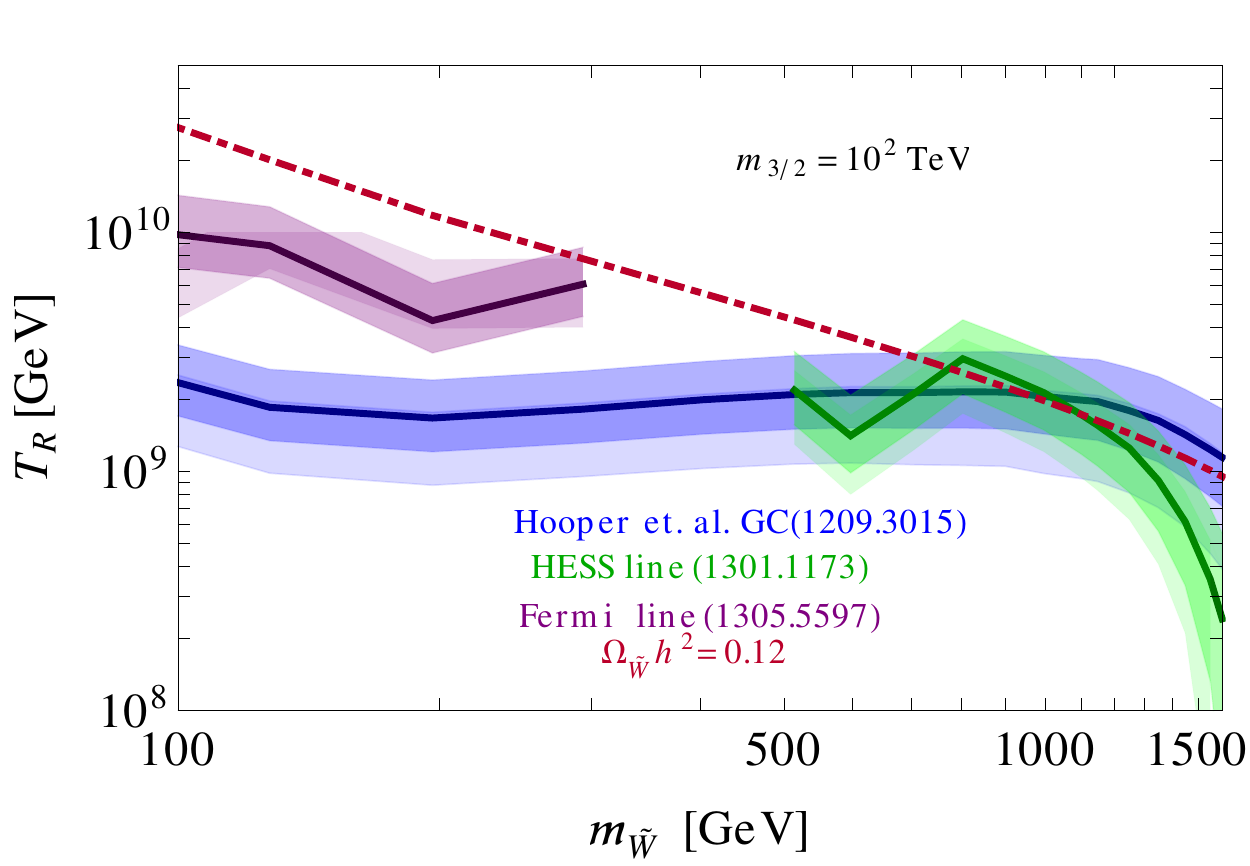}\includegraphics[width=0.5\textwidth]{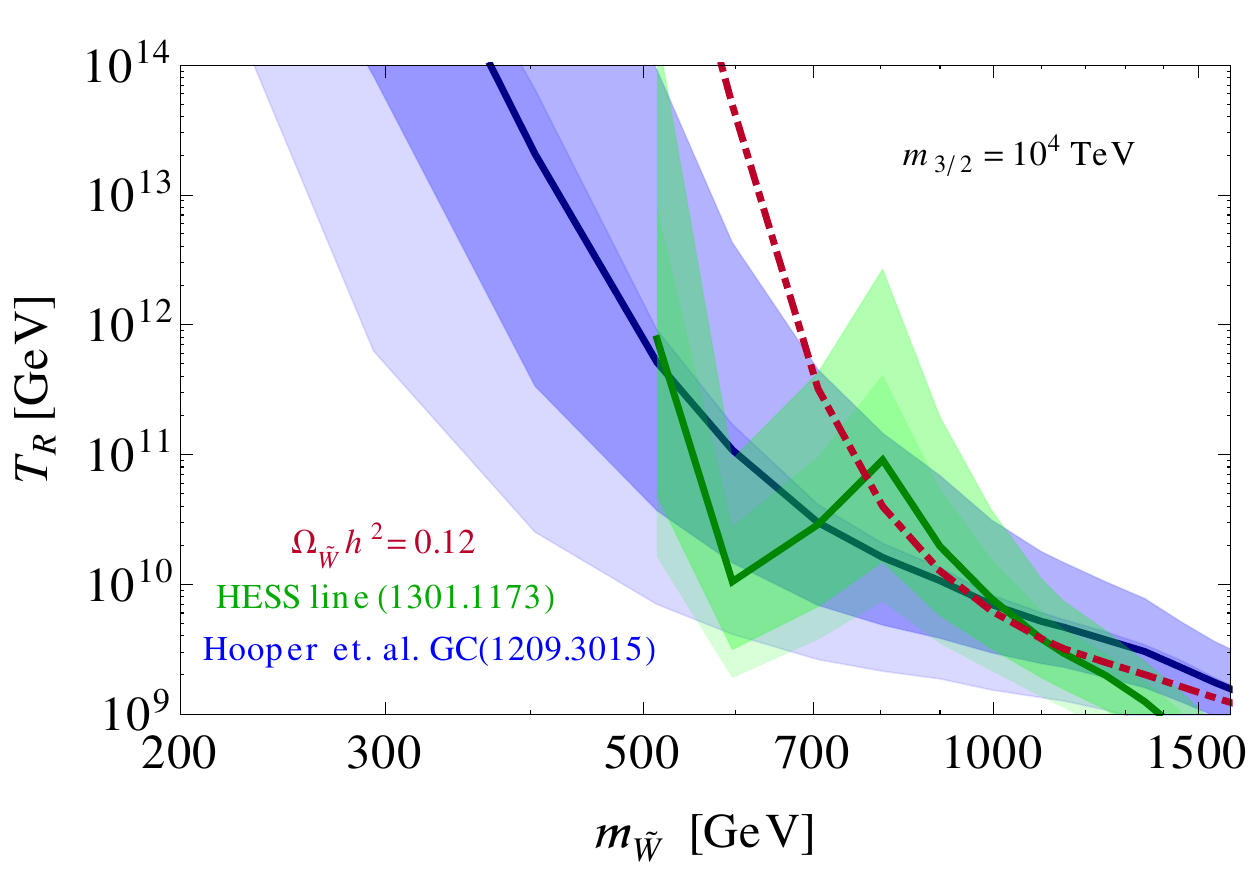}
\end{center}
\caption{\small Upper bounds on inflaton reheating temperature $T_{R}$ as a function of wino mass for $m_{3/2} = 100$ TeV (left) and $m_{3/2} = 10^4$ TeV (right). The blue, purple, green curves with bands around them correspond to constraints from Fermi galactic center continuum, Fermi line search and HESS line search respectively. The bands are derived by varying parameters of NFW (Einasto) dark matter profiles in the 2$\sigma$ range~\cite{Iocco:2011jz}. The burgundy dot-dashed line corresponds to the upper bound derived from requiring $\Omega_{\tilde{W}} h^2 = 0.12$.}
\label{fig:reheating}
\end{figure}

We express the constraints on allowed non-thermal $\Omega_{{\tilde{W}}} h^2$ as an upper bound on the inflaton reheating temperature $T_R$ as a function of wino mass for $m_{3/2} = 100$ TeV and $10^4$ TeV in Fig.~\ref{fig:reheating}. In this figure, we assumed that freeze-in contribution to the primordial gravitino relic abundance is negligible. As mentioned at the end of last section, taking into account of the freeze-in contribution will only make the upper bound stronger. 
  
The left panel of Fig.~\ref{fig:reheating} stays almost unmodified for 10 TeV $< m_{3/2} < 10^4$ TeV as the wino annihilation is ineffective and the relic abundance is independent of $m_{3/2}$ as can be seen from the first term in Eq.~\eqref{no-annOmW}. The reheating temperature is bounded to be below $3 \times 10^9$ GeV for the whole wino mass range. For wino mass close to 1.5 TeV, the HESS constraint pushes the reheating temperature to be even lower to about a few times $10^8$ GeV.

In the right panel of Fig.~\ref{fig:reheating}, the gravitino mass is set to be $10^4$ TeV. In this case, for light wino with mass below 300 GeV, wino annihilation becomes effective and its relic abundance is insensitive to the reheating temperature as shown in Eq.~\eqref{ann-OmW}.  
Therefore, the upper bound on the inflaton reheating temperature is lifted up entirely. For heavier wino, the annihilation rate drops with the increasing mass and the wino relic abundance interpolates between Eq.~(\ref{ann-OmW}) and Eq.~(\ref{no-annOmW}). In the whole wino mass range, the upper bound on $T_R$ is above $10^9$ GeV. For even heavier gravitino, the bound on $T_R$ becomes even weaker.

\begin{figure}[!h]\begin{center}
\includegraphics[width=0.5\hsize ,height= 6.0cm]{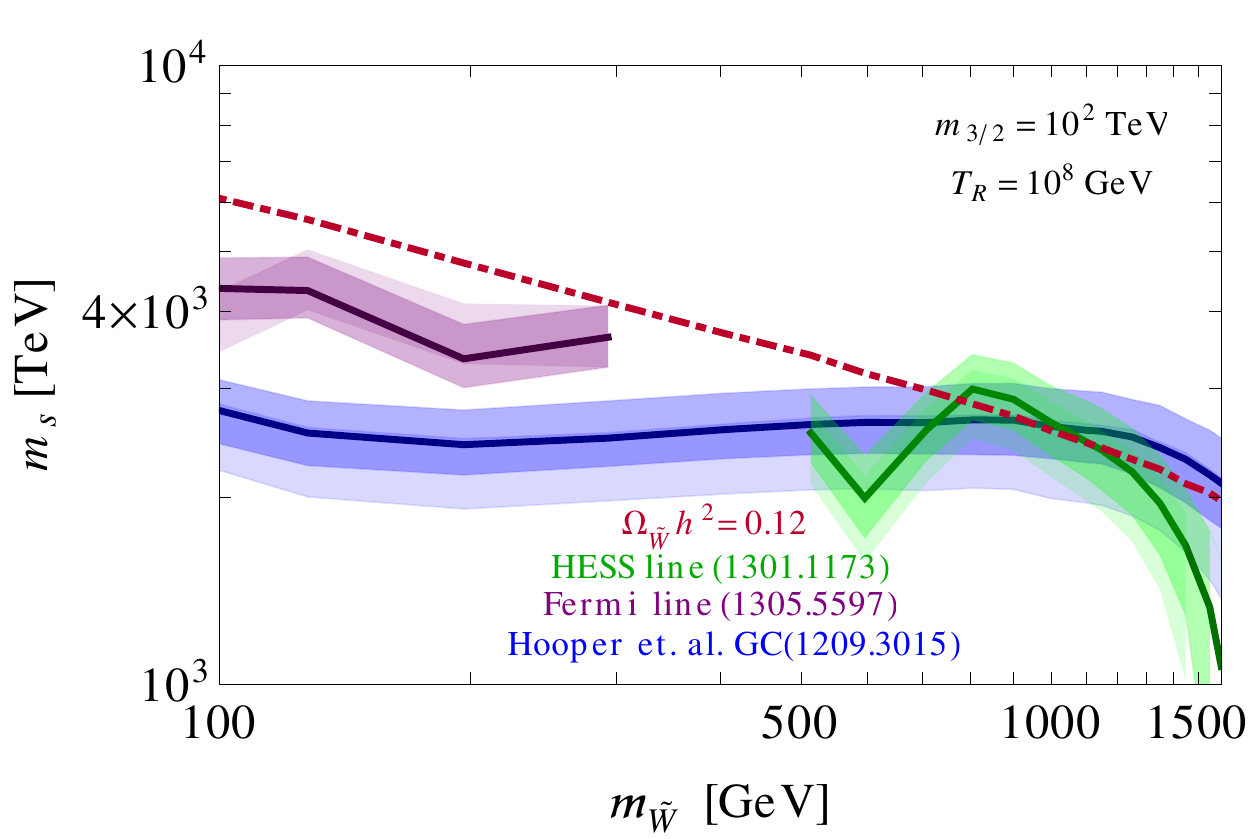}\includegraphics[width=0.5\hsize ,height=6.0cm]{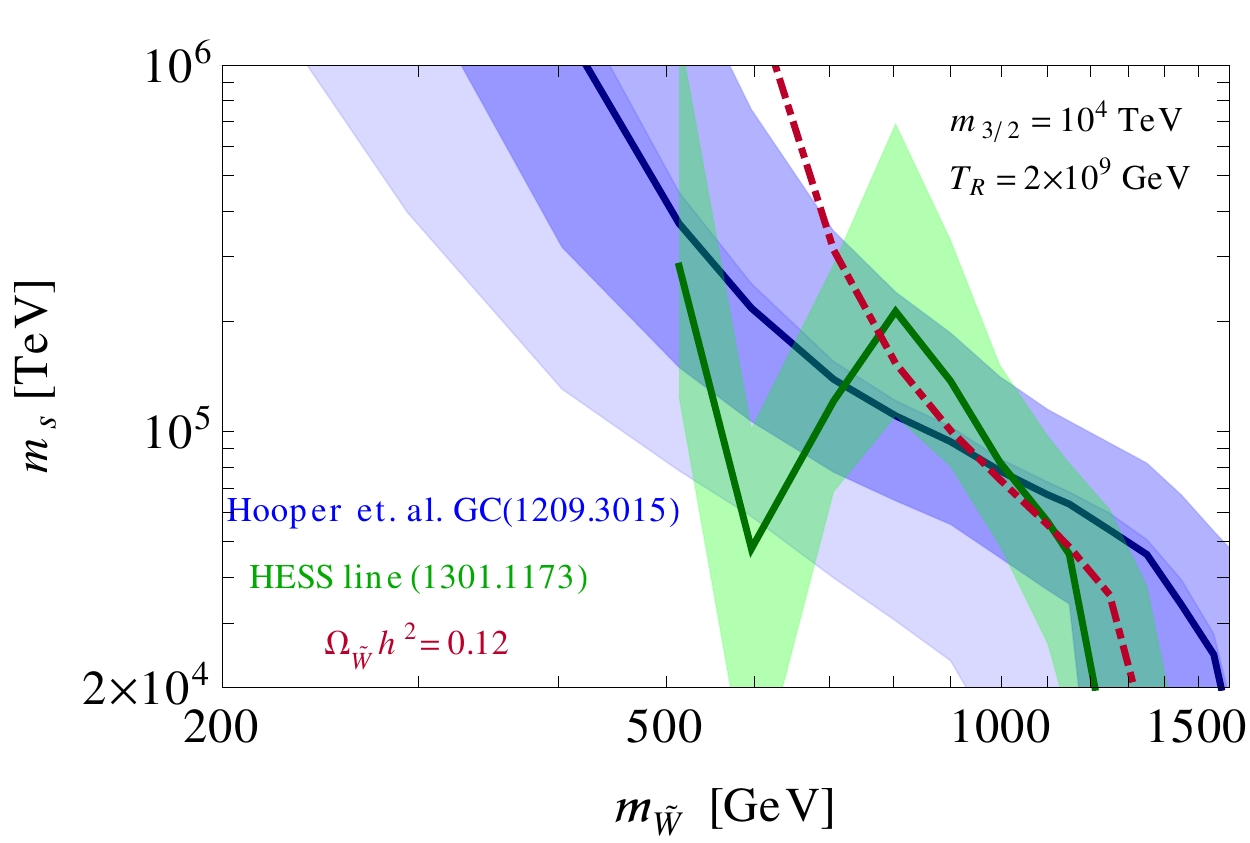}
\end{center}
\caption{\small Upper bounds on scalar mass  $m_s$ as a function of wino mass for $m_{3/2} = 100 ~\rm{TeV} $(left) and $m_{3/2} = 10^4 ~\rm{TeV} $ (right). The blue, purple, green curves with bands around them correspond to constraints from Fermi galactic center continuum, Fermi line search and HESS line search respectively. The bands are derived by varying parameters of NFW (Einasto) dark matter profiles in the 2$\sigma$ range~\cite{Iocco:2011jz}. The burgundy dot-dashed line corresponds to the upper bound derived from requiring $\Omega_{\tilde{W}} h^2 = 0.12$. }
\label{fig:mscalar}
\end{figure}

One could also consider upper bound on the SUSY scalar masses, $m_s$, which is depicted in Fig.~\ref{fig:mscalar}. In the left panel, we took $m_{3/2} = 100$ TeV and $T_R=10^8~\rm{GeV}$ so that the thermal scattering contribution is negligible. Increasing the reheating temperature will only make the bound even stronger. In this case, indirect detection constraints restrict the scalar mass to be below $(2 - 3) \times 10^3$ TeV for the whole wino mass range. In the right panel, we set $m_{3/2} = 10^4$ TeV and $T_R = 2 \times 10^9$ GeV. Since this is the case where wino annihilation becomes more effective, the upper bounds on the SUSY scalar mass depends less on $T_R$ and are reduced significantly compared to the case with a lighter gravitino. More specifically, for wino above 300 GeV, indirect detection constraints restrict scalar masses to be below $10^4 - 10^6$ TeV. For wino below 300 GeV, the upper bound is almost lifted up entirely. 

\section{Implications of the BICEP2 Result}
\label{sec:bicep}
Recently the BICEP2 collaboration reported a groundbreaking discovery of inflationary gravitational waves in the $B$-mode power spectrum in the range $30 < l < 150$~\cite{Ade:2014xna}. The observed $B$-mode spectrum is well fit by a lensed $\Lambda$-CDM plus tensor model with a tensor-to-scalar ratio $r=0.20^{+0.07}_{-0.05}$. Such a large tensor-to-scalar ratio has a profound implication for the inflation paradigm. Notice that if running of the spectral index is allowed, the combined Planck and BICEP data could have a different best fit. In our paper, we will not explore this possibility as we don't expect $r$ to change much. We will first review the basics of tensor-to-scalar ratio in the slow-roll inflation paradigm for completeness in Sec.~\ref{sec:basics}. Readers who are familiar with this topic could skip this section. Then we will discuss the implications of BICEP2 result for the inflation mass scale and reheating temperature in Sec.~\ref{sec:reheating}. 

\subsection{Basics of Tensor-to-Scalar Ratio}
\label{sec:basics}
We will follow closely Lecture 2 in Ref.~\cite{Baumann:2009ds} in this brief review. In slow-roll inflation models, the metric perturbation during the inflation period could be decomposed into scalar and tensor modes, which result in density and gravitational wave fluctuations respectively. Each mode could be characterized by a fluctuation amplitude squared~\cite{Mukhanov:1985rz, Mukhanov:1988jd} 
\beq
\Delta_s^2(k) &=& \frac{H^4}{4\pi^2 \dot{\phi}^2} = \frac{1}{12\pi^2M_p^6}\frac{V^3}{V^{'2}} \quad {\rm scalar} \\
\Delta_t^2(k) &=& \frac{2H^2}{\pi^2 M_p^2} = \frac{2}{3\pi^2} \frac{V}{M_p^4}, \quad {\rm tensor}
\eeq
where the reduced Planck scale is $M_p = 2.4 \times 10^{18}$ GeV. It should be understood that all the physical quantities above are evaluated at horizon crossing $k=aH$ at which the relevant comoving scales for the CMB exits the Hubble radius. $\dot{\phi}$ is the time derivative of the inflaton field $\phi$ and $V^{'}$ is the derivative of the inflaton potential with respect to $\phi$. In deriving the second expression of the amplitude squared in each line, we used equation of motion for the inflaton $H^2\approx V/(3M_p^2)$ and $\dot{\phi} \approx - V^{'}/(3H)$.\footnote{One easy way to understand the appearance of $\dot{\phi}^2$ in the scalar perturbation amplitude squared is through effective field theory (EFT)~\cite{Cheung:2007st}. The key insight of inflation EFT is that the inflaton spontaneously breaks time translation invariance and results in a Goldstone mode ``eaten" by the graviton to appear in the scalar modes. Compared to the tensor mode, the kinetic term for the Goldstone (scalar) mode has an additional factor of $\dot{H}$ in the kinetic term, which signals the break down of EFT in the limit of pure de Sitter space $\dot{H} = 0$. By equation of motion, $\dot{H}$ is proportional to $\dot{\phi}^2$ and consequently $\dot{\phi}^2$ appear in the denominator of scalar fluctuation amplitude squared. } 

Normalizing the scalar spectrum to the COBE~\cite{Smoot:1992td} or WMAP~\cite{Spergel:2006hy} anisotropy measurement gives $\Delta_s^2(k) \approx 2.2 \times 10^{-9}$. Then one could define tensor-to-scalar ratio $r \equiv \Delta_t^2(k)/\Delta_s^2(k)$, which directly measures the inflation energy scale 
\beq \label{eq:potentialenergy}
V\approx (1.8  \times 10^{16} \, {\rm GeV})^4 \left( \frac{r}{0.1}\right). 
\eeq
$r$ also relates directly to the evolution of the inflaton as
\beq
r=\frac{8}{M_p^2} \left(\frac{d \phi}{d N}\right)^2,
\eeq
where differential $e$-folds $dN = H dt$. Then one could write the field displacement between the time when CMB fluctuations exited the horizon at $N_{\rm cmb}$ and the end of inflation at $N_{\rm end}$ in terms of an integral
\beq \label{eq:integral}
\frac{\Delta \phi}{M_p} = \int^{N_{\rm cmb}}_{N_{\rm end}} dN \sqrt{\frac{r}{8}}.
\eeq
Setting $N_{\rm end} = 0$ and given that $N_{\rm cmb} \approx (40 - 60)$ and $r$ is approximately constant during the inflation era, one obtains the famous Lyth bound~\cite{Lyth:1996im}
\beq \label{eq:displacement}
\frac{\Delta \phi}{M_p} \approx 6.7 \left(\frac{N_{\rm cmb}}{60}\right)  \sqrt{\frac{r}{0.1}}. 
\eeq

Inspecting Eq.~(\ref{eq:potentialenergy}) and (\ref{eq:displacement}), one could see immediately that the BICEP2 result points towards a large field displacement of order Planck scale during inflation or in other words, large field inflation. Existing examples of large-field inflation include chaotic inflation where a single power term dominates the potential~\cite{Linde:1986fc, Linde:1986fd} 
\beq
V(\phi) = \lambda_p \phi^p, 
\eeq
and natural inflation with a periodic potential resulting from a shift symmetry the inflaton enjoys~\cite{Freese:1990rb}
\beq
V(\phi) = V_0 \left(1+\cos \left(\frac{\phi}{f}\right)\right).
\eeq

\subsection{Implication for Reheating Temperature}
\label{sec:reheating}
Now we want to estimate the inflaton mass scale. We start with a toy model of large field inflation $V=m_\phi^2\phi^2$. In this model, the scalar fluctuation amplitude squared is 
\beq
\Delta_s^2(k) = \frac{m_\phi^2}{M_p^2} \frac{N_{\rm cmb}^2}{3\pi^2},
\eeq
where $N_{\rm cmb} =\phi_{\rm cmb}^2/(4M_p^2)$. 
Given the normalization to the CMB measurement, $\Delta_s^2(k) \approx 2.2 \times 10^{-9}$, the inflaton mass is 
\beq
m_\phi \approx 10^{13} \,{\rm GeV} \, \left(\frac{60}{N_{\rm cmb}}\right)^2.
\eeq
One could check in more realistic models such as chaotic inflation and natural inflation that the inflaton mass scale is around $10^{13}$ GeV~\cite{Harigaya:2014sua, Harigaya:2014qza, Freese:2014nla}. One crude estimate of the inflaton mass in all these large-field inflation model is 
\beq
m_\phi^2 \sim \frac{V}{(\Delta \phi)^2} \approx \left(2 \times 10^{13} \, {\rm GeV}\right)^2,
\eeq
where we used Eq.~(\ref{eq:potentialenergy}) and~(\ref{eq:displacement}) assuming $N_{\rm cmb} = 60$. 

After inflation ends, inflaton starts to oscillate around the minimal of the potential. Its coupling to other particles induce conversion of the inflationary energy into the SM degrees of freedom. The reheating temperature is then determined by the inflaton decay width $\Gamma_\phi$ as 
\beq
T_R = \left(\frac{10}{g_*(T_R) \pi^2}\right)^{1/4}\sqrt{\Gamma_\phi M_p} \approx 0.3 \sqrt{\Gamma_\phi M_p},
\eeq
where we took $g_*(T_R) \approx 200$. 
The simplest possibility is that inflatons decay through renormalizable couplings to lighter degrees of freedom. For example, the decay width is $\Gamma_\phi = y^2 m_\phi/(8\pi)$ for inflaton coupling to fermions with a Yukawa coupling $y$. Then the reheating temperature is
\beq
T_R \approx 3 \times 10^{11} \, {\rm GeV} \left( \frac{y}{10^{-3}}\right) \sqrt{\frac{m_\phi}{10^{13}\,{\rm GeV}}}.
\eeq
Notice that Yukawa coupling larger than $10^{-5}$ only makes sense in supersymmetric scenarios where the one-loop quantum correction does not modify the inflaton potential much due to a cancelation between fermionic and bosonic contributions. 

If the renormalizable couplings of inflaton to lighter particles are negligible (e.g., $y < 10^{-5}$), it would always decay through Planck-scale suppressed operators. At the leading order, the inflaton decay width and the corresponding reheating temperature are
\beq
\Gamma_\phi = \frac{c m_\phi^3}{M_p^2}, \quad T_R\approx 5\times 10^9 \, {\rm GeV} \sqrt{c} \left(\frac{m_\phi}{10^{13}\,{\rm GeV}}\right)^{3/2},
\eeq
where $c$ is some order one number determined by quantum gravity. From the point of view of operator analysis, this decay is induced by dimension five operators such as $\phi F\tilde{F}/M_p$ with $F$ the field strength of SM gauge interaction.  
In other words, the BICEP2 results imply a minimal reheating temperature at or above $10^9$ GeV! 

One should worry about the caveats of the very simple estimate above. One question is whether the leading order gravitational couplings through dimension five operators could be suppressed and the reheating temperature could be even lower. This could be true if the inflaton is charged under a gauge symmetry (global symmetry is not respected by quantum gravity) and then dimension five operators are forbidden. This is an interesting possibility but we will not explore it here further but leave it for future work. Another concern is that since reheating is a very complicated process (for a review, see~\cite{Bassett:2005xm}), our simple estimate of a minimal reheating temperature might be misleading. In particular, there could exist a preheating era in which particles coupled to the inflaton are resonantly produced by parametric resonance and the temperature of the plasma could be higher than the reheating temperature. Yet preheating might make the tension between the upper bound on $T_R$ derived in Sec.~\ref{sec:relic} and Sec.~\ref{sec:indirect} and the lower bound on $T_R$ derived in this section even worse. The reason is that gravitinos could be over-produced non-thermally during the preheating era~\cite{Kallosh:1999jj, Giudice:1999am, Kallosh:2000ve, Maroto:1999ch,Nilles:2001fg}.\footnote{In certain supergravity models, the non-thermal production could be suppressed~\cite{Nilles:2001ry, Greene:2002ku}.} 
Nonetheless, it is interesting and important to carry out a thorough study of preheating/reheating in sound (stringy) inflation models.

\section{Implications for SUSY}
\label{sec:implications}
So far we have demonstrated a (mild) tension between mini-split SUSY with a heavy unstable gravitino at around the PeV scale and the BICEP2 result. In mini-split SUSY, the reheating temperature has to be below $10^9$ GeV to avoid overproduction of DM particles from gravitino decay while the BICEP2 result prefers large-field inflation with a reheating temperature above $10^9$ GeV. In other words, the BICEP2 results favor a larger splitting between the gravitino and the gauginos than the one-loop factor if the gauginos are fixed at around the TeV scale. Interestingly, \textbf{the requirement that gaugino mass does not exceed the TeV scale constrains the gravitino mass to be around or below $10^{13}$ GeV, which is also the mass scale of the inflaton implied by BICEP2 !} Below we will review the derivation of this statement by operator analysis following Refs.~\cite{ArkaniHamed:2004fb,ArkaniHamed:2004yi}.  

In supergravity, the easiest way to cancel the positive vacuum energy from SUSY breaking contribution is to have a non-zero VEV of the superpotential. As the superpotential $W$ carries $R$-charge 2, its VEV $W_0$ breaks $U(1)_R$ symmetry spontaneously. It also gives a gravitino mass 
\beq
m_{3/2} \approx \frac{W_0}{M_p^2} \approx \frac{\left|F_X\right|}{\sqrt{3} M_p},
\eeq
where $F_X$ is the $F$-term VEV of the SUSY breaking spurion $X$. 
A non-zero gaugino mass is generated only when both $U(1)_R$ and SUSY are broken. The lowest dimensional operator built out of SUSY breaking spurion $X$, $U(1)_R$ breaking spurion $W$ and MSSM superfields arise in the K{\"a}hler potential
\beq
\int d^2 \theta d^2 \bar{\theta} \frac{X^\dagger X W W_\alpha W^\alpha }{M_*^6},
\eeq
where $W_\alpha$ denotes the MSSM gauge supermultiplet. This operator could be generated by gravitational loops where $M_* \sim M_p$ and gives a minimal contribution to the gaugino mass 
\beq
m_{1/2} \gtrsim \frac{\left|F_X\right|^2W_0}{M_p^6} \approx \frac{3m_{3/2}^3}{M_p^2}.
\eeq
Requiring $m_{1/2}$ at or below TeV leads to $m_{3/2} \lesssim 10^{13}$ GeV! This large hierarchy between gravitino and gaugino could be realized in no-scale supergravity which could arise from the Scherk-Schwarz mechanism~\cite{Ellis:1984bm, Scherk:1979zr}.

\section{Conclusions and Outlook}
\label{sec:con}
In this paper, we study the implication of DM indirect detection and BICEP2 in the split SUSY scenario with a heavy unstable gravitino. In the mini-split spectrum with scalars/gravitinos only one-loop factor above the TeV-scale gauginos, the reheating temperature has to be low to avoid overproduction of DM particles from gravitino decays. In particular, we demonstrate that indirect detection requires the reheating temperature to be below about $10^9$ GeV if the wino is (a component of) DM. On the other hand, the large tensor-to-scalar ratio observed by BICEP2 favors large-field-inflation with a reheating temperature around or above $10^9$ GeV. Given this mild tension and the phenomenological upper bound on the gravitino mass derived by requiring the gauginos to be at the TeV scale, it is tempting to think more seriously of the (highly) split SUSY scenario in which inflaton/gravitino are at around $10^{13}$ TeV and gauginos are still at the TeV scale with lightest neutralino being (part of) DM.\footnote{Axion could be the dominant DM component.} Indeed this picture has recently been discussed in the framework of Intermediate Scale SUSY~\cite{Hall:2014vga}.  

In general, given the BICEP2 result, it is very interesting to use the scale of inflation to probe the full range of split SUSY scenarios through observables such as equilateral non-gaussianity~\cite{Craig:2014rta}. It will also be of interest to study the implications of the BICEP2 result for baryogenesis. For example, thermal leptogenesis works for a reheating temperature above $2\times 10^9$ GeV~\cite{Giudice:2003jh}, which fits well with the BICEP2 result.

\section*{Acknowledgements}
We thank Haipeng An, Joseph Bramante, Tim Cohen, Raffaele Tito D'Agnolo, Matt Reece, Kuver Sinha and Scott Watson for useful comments. JF thanks the Center for Future High Energy Physics in Beijing for hospitality where part of the project is carried out. JF does not thank the thief who stole her laptop before she could save the draft to dropbox in the final stage of the project.  

\section*{Appendix A: Gravitino from Inflaton Decay}

In this appendix, we review non-thermal gravitino production from inflaton decays. In general, decays of inflaton can overproduce gravitinos which subsequent decays can induce LSP overproduction~\cite{Kawasaki:2006gs,Kawasaki:2006hm}. 
Consider the following simple model of SUSY breaking and inflation~\cite{Nakayama:2012hy},
\beq
\label{Kahler}K &=& |\phi|^2 + |X|^2 + |z|^2-\frac{|z|^4}{\tilde{\Lambda}^2},\\
W &=& X\left(g\frac{\phi^n}{M_{p}^{n-2}} - v^2\right)+\mu^2 z + W_0,
\eeq
where $z$ is the SUSY breaking spurion and $\tilde{\Lambda}$ is the QCD scale of the dynamical SUSY breaking sector. Here, $\mu$ is the SUSY breaking scale related to the $F$-term VEV of $z$ through $F_z\simeq-\mu^2\simeq \sqrt{3}m_{3/2}M_{p}$ and $W_0 $ is the constant term introduced in order to cancel the positive vacuum energy from SUSY breaking in order to obtain a vanishing cosmological constant. 

The scalar potential in supergravity is given by
\beq\label{SP}
V = e^{K/M_{pl}^2}\left[K_{i\bar{j}}^{-1}(D_i W)(D_{\bar{j}}W)-3\frac{|W|^2}{M_{p}^2}\right],
\eeq
where $D_i$ is the covariant derivative with respect to field $i$. 
There is a mass mixing between $X$ and $z$ arising from the following terms in the scalar potential above \eqref{SP},
\beq
V \supset \left| \frac{X n g \phi^{n-1}}{M_{p}^{n-2}} + \frac{\phi^{\dagger} W}{M_{p}^2}              \right|^2 \approx  \frac{m_\phi \langle \phi \rangle \mu^2}{M_{p}^2} X z^{\dagger} + h.c,
\eeq
where $m_\phi = n g\langle \phi \rangle^{n-1}/M_p^{n-2}$. 

The operator $|z|^4/\tilde{\Lambda}$ in the K{\"a}hler potential induces $z$ decaying into the goldstino pair $(\tilde{z})$ via
\beq
\mathcal{L} \supset \int d^2\theta d^2\bar{\theta} K \sim -2 \frac{F_z^{\dagger}}{\tilde{\Lambda}^2} z^{\dagger} \tilde{z}\tilde{z} + h.c,
\eeq 
where the decay rate is given by 
\beq
\Gamma_{z\to \tilde{z}\tilde{z}} \simeq \frac{1}{96\pi}\frac{m_z^5}{m_{3/2}^2M_{p}^2}.
\eeq
Since the goldstino is ``eaten'' by the gravitino via the super-Higgs mechanism, the decay rate above can be expressed as the decay rate of the inflaton into a pair of gravitinos via the mass mixing with $z$:
\beq\label{infto3/2}
\nn \Gamma_{\phi\to \tilde{z}\tilde{z}} &\sim & \left(\frac{\theta}{\sqrt{2}}\right)^2 \frac{m_\phi}{m_z}~ \Gamma_{z\to \tilde{z}\tilde{z}},\\
&\sim & \left(\frac{\theta}{\sqrt{2}}\right)^2 \left(\frac{m_z}{m_\phi}\right)^4 \frac{m_{\phi}^5}{m_{3/2}^2 M_{p}^2} 
\eeq 
where the mixing angle between inflaton $\phi$ and $z$, $\theta$, is given by $\sqrt{3}(m_{3/2}\langle \phi \rangle)/(m_\phi M_{p})$ for $m_{\phi}\gg m_{z}$, $\sqrt{3}(m_{3/2}m_\phi \langle \phi \rangle)/(m_z^2 M_{p})$ for $m_\phi\ll m_z$. Therefore, in the case that $m_{\phi}\gg m_{z}$, the decay rate \eqref{infto3/2} of inflaton into a pair of gravitino is suppressed by $(m_z/m_\phi)^4$. 

If $z$ is only charged under some global symmetry, one could not forbid operators such as $|\phi|^2 z$ and $|\phi|^2 zz$ in the K{\"a}hler potential \eqref{Kahler}. These operators will always be induced by Planck scale physics as it only respects local symmetries \cite{Kamionkowski:1992mf}. These operators are dangerous as they would enhance the decay rate of inflaton to gravitinos by $m_\phi^2/m_{3/2}^2$. Thus in addition to the hierarchy $m_{3/2} \ll m_z \ll m_\phi$, the SUSY breaking spurion cannot be a gauge singlet!

\bibliography{ref}
\bibliographystyle{utphys}
\end{document}